\documentclass[fleqn,twoside]{article}
\usepackage{espcrc2}

\usepackage{graphicx}

\newcommand{\AmS}{{\protect\the\textfont2
  A\kern-.1667em\lower.5ex\hbox{M}\kern-.125emS}}

\title{Study of hyperfine structure in simple atoms and precision tests of the bound state QED}

\author{S. G. Karshenboim\address[MPQ]{Max-Planck-Institut f\"ur Quantenoptik,
85748 Garching, Germany}\address[VNIIM]{D. I. Mendeleev Institute
for Metrology (VNIIM),  St. Petersburg 190005, Russia}, S. I.
Eidelman\address{Budker Institute for Nuclear Physics and
Novosibirsk State University, Novosibirsk, 630090, Russia}, P.
Fendel\addressmark[MPQ], V. G. Ivanov\address{Pulkovo Observatory,
St. Petersburg 196140, Russia}, N. N. Kolachevsky\address{P. N.
Lebedev Physics Institute, Moscow, 119991, Russia}, V. A.
Shelyuto\addressmark[VNIIM], and
T.~W.~H\"{a}nsch\addressmark[MPQ]}

\begin{document}

\begin{abstract}
We consider the most accurate tests of bound state QED, precision
theory of simple atoms, related to the hyperfine splitting in
light hydrogen-like atoms. We discuss the HFS interval of the $1s$
state in muonium and positronium and of the $2s$ state in
hydrogen, deuterium and helium-3 ion. We summarize their QED
theory and pay attention to involved effects of strong
interactions. We also consider recent optical measurements of the
$2s$ HFS interval in hydrogen and deuterium. \vspace{1pc}
\end{abstract}

\maketitle

\section{Introduction}

Light simple atoms are basically described by quantum
electromagnetic theory. Quantum electrodynamics (QED) is well
established and in particular it covers all interactions of
leptons (electrons and muons) and photons. Such a lepton-photon
theory is obviously incomplete because even pure leptonic systems
are not free of hadronic effects which enter through virtual
hadronic intermediate states. Effects of strong interactions
cannot be calculated {\em ab initio\/} and additional experimental
data and/or phenomenological models are needed. Here we consider
QED tests with hyperfine splitting in light hydrogen-like atoms
paying attention to both: basic {\em ab initio\/} QED theory and
relatively small, but most uncertain, hadronic contributions.

An application of QED to the bound state problem, bound state QED, is
much more complicated than ordinary QED and it deserves serious
tests. Some of such tests are significant for the determination of
fundamental constants and in particular of the fine structure
constant $\alpha$, which may be obtained from the hyperfine
structure (HFS) interval in muonium (see reviews
\cite{PRep1,PRep2,codata} for more detail).

The HFS interval in hydrogen and some other light atoms has been
known for a while with an experimental accuracy at the level of a
part in $10^{12}$. Meanwhile, the related theory suffers from
uncertainties of the nuclear structure effects at one-ppm level.

Here, we consider a few possibilities to perform QED tests going
far beyond this level of accuracy.

\section{Studying the $1s$ hyperfine splitting}

\subsection{The $1s$ hyperfine interval in muonium}

Problems in accurate calculations of the proton or nuclear
structure effects drew attention to studies of pure leptonic atoms
such as a bound system of a positive muon and an electron, the
muonium. In contrast to the hydrogen atom, the nucleus, a muon, is
free of effects of strong interactions. Nevertheless, those
effects enter through hadronic vacuum polarization. That sets an
ultimate limit on any QED tests with muonium. Uncertainties of the
QED theory and of calculations of the hadronic effects are
presented in \cite{strong,PRep1}. Muonium is of metrological
interest due to determination of the fine structure $\alpha$,
muon-to-electron mass ratio $m_\mu/m_e$ and some other fundamental
constants \cite{codata}.

A calculation of the hadronic effects \cite{strong} is similar to
those for the anomalous magnetic moment of the muon \cite{muon}.
It is based on the low energy $e^+e-$ data which accuracy,
although fast improving, is still behind the measurement of the
muon $g\!-2\!2$ value \cite{muonexp}. The current difference
between the experimental value and its theoretical prediction
differs from zero by almost three standard
deviations \cite{muonth}. Attempts to increase the accuracy
of the prediction by adding data on the $\tau$ lepton decays
revealed one more possible deviation from the Standard Model
expectation \cite{muonth}.

\begin{table*}[htb]
\caption{Theory of the specific difference $D_{21}=8E_{\rm
HFS}(2s)-E_{\rm HFS}(1s)$ in light hydrogen-like atoms. The
numerical results are presented for the related frequency
$D_{21}/h$. QED3 and QED4 stands for the third and fourth order
QED corrections in units of the so-called Fermi energy.
\label{T:th}} \label{table:1}
\newcommand{\m}{\hphantom{$-$}}
\newcommand{\cc}[1]{\multicolumn{1}{c}{#1}}
\renewcommand{\tabcolsep}{1.5pc} 
\renewcommand{\arraystretch}{1.2} 
\begin{tabular}{@{}lllll}
\hline
Contribution to HFS in & Hydrogen, [kHz] & Deuterium, [kHz] & $^3$He$^+$ ion, [kHz]\\
\hline
$D_{21}({\rm QED3})$ & 48.937 &  11.305\,6 & -1\,189.253\\
$D_{21}({\rm QED4})$ & {0.018(5)} & {0.004\,4(10)}  &-1.13(14)\\
$D_{21}({\rm Nucl})$ & {-0.002} & {0.002\,6(2)} & 0.307(35)\\
\hline
$D_{21}({\rm theo})$ & 48.953(5) &  11.312\,5(10) & -1\,190.08(15)  \\
\hline
\end{tabular}\\[2pt]
\end{table*}

\subsection{The $1s$ hyperfine interval in positronium}

Another pure leptonic atomic system is positronium. The nucleus, a
positron, is a light one. That means that various recoil effects,
which are crucially important in any advanced QED theory of the
hyperfine splitting, are enhanced. As a consequence,  critical QED
tests performed on positronium can be competitive with other
tests (such as muonium experiments) even at a
relatively low experimental accuracy (see, e.g., \cite{PRep1}).

Enhancement of the recoil effects has one more consequence. In
conventional atoms (such as hydrogen), theory of the
spin-independent energy shifts (the Lamb shift) is completely
different from that for the HFS effects. In the Lamb shift theory
and the theory of the $1s-2s$ transition, higher-order two-loop
external-field effects dominate in the uncertainty budget, while
the recoil and the hyperfine effects are responsible for
relatively small corrections. In the theory of the HFS interval,
the recoil effects are most important for the uncertainty, while
the external field effects are under control.

For positronium, the theory of the $1s-2s$ transition is in a sense
similar to the HFS theory and dominant uncertainty comes from the
HFS effects (see, e.g., \cite{PRep1}). However, the related
accuracy is somewhat lower than that of the $1s$ HFS interval.

\section{The $2s$ hyperfine interval: \protect\newline Theory of the
specific difference $D_{21}$}

Another accurate QED test is possible with ordinary light
hydrogen-like atoms. One can combine the HFS intervals of the $1s$
and $2s$ states in the same atom
\begin{equation}
D_{21} = {2^3}\cdot E_{\rm HFS}(2s)-E_{\rm HFS}(1s)
\end{equation}
to eliminate the leading nuclear contributions (see., e.g.,
\cite{th2s}).

A substantial cancellation of the nuclear structure contributions
takes place because the nuclear contributions in the leading
approximation are of the factorized form
\begin{equation}
\Delta E({\rm Nucl}) = {A({\rm Nucl})} \times {\big\vert\Psi_{nl}
({\bf r}=0)\big\vert^2}\;,
\end{equation}
i.e., the correction is a product of the nuclear-structure
parameter {$A({\rm Nucl})$} and the wave function at the origin
\begin{equation}
{ \big\vert\Psi_{nl}({\bf r}=0)\big\vert^2 = \frac{1}{\pi}
\left(\frac{Z\alpha\,m_R c}{ {n} \hbar} \right)^{{3}} }\delta_{l0}
\;,
\end{equation}
where $n,l$ are the principal and orbital quantum numbers,
respectively, and $m_R$ is the reduced mass.

Higher-order corrections due to nuclear effects are of a more
complicated form and some of them survive this cancellation.
Still, they are much smaller and under control \cite{th2s}. The
theory of $D_{21}$ in light hydrogen-like atoms is presented in
Table~\ref{T:th} \cite{th2s,th2s1}.

The $1s$ and $2s$ hyperfine intervals have been measured much more
accurately compared to the theoretical prediction which can be
made for each of them separately. Meanwhile, for the difference
$D_{21}$ the experimental and theoretical accuracy are
competitive. While for $^3$He$^+$ the experiment \cite{He2s} is
still somewhat more accurate than theory \cite{th2s1}, in the case
of hydrogen and deuterium, the theory is more accurate than the
measurement of the HFS interval in the $2s$ state.

\section{The $2s$ hyperfine interval: optically measured in
hydrogen and deuterium}

Measurements of the $2s$ HFS interval in hydrogen \cite{H2s_old}
and deuterium \cite{D2s_old} by microwave means have nearly a
fifty year history. The hydrogen result was somewhat improved in
2000 \cite{H2s_new} by traditional microwave means.

Recently a new generation of optical experiments was launched
using a hydrogen spectrometer developed at Max-Planck-Institut
f\"ur Quantenoptik for the ultraviolet $1s-2s$ transition
\cite{1s2s}. The spectrometer was developed in order to build a
natural frequency standard.

\begin{figure}[htb]
\centerline{\includegraphics[scale=0.40]{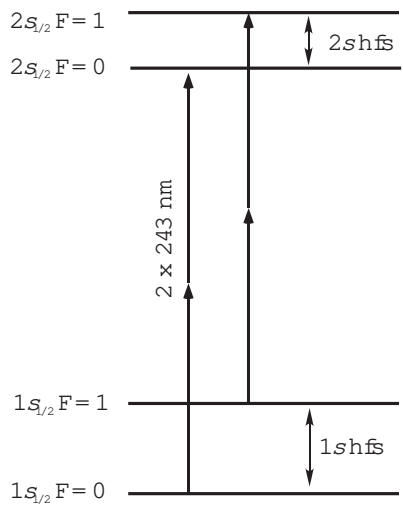}}
\caption{The $1s$ and $2s$ levels in the hydrogen atom. Not to
scale.} \label{fig:opticalhfs}
\end{figure}

The fractional uncertainty of the former measurements \cite{1s2s}
was at the level of a few parts in $10^{14}$ or 30--40 Hz and was
due to various systematic effects. Expecting that dominant
systematic effects are spin-independent, one can hope that a
comparison of spin components of the $1s-2s$ line can be performed
with a higher absolute accuracy. The $1s-2s$ transition lies in
the ultraviolet domain (see Fig.~1) and for the triplet case the
result is $2\,466\,061\,102\,474\,851(34)$ Hz \cite{1s2s}.

We have compared the $1s-2s$ ultraviolet frequencies for different
HFS components in hydrogen \cite{H2s} and deuterium \cite{D2s} and
with the value of the $1s$ HFS intervals known for both atoms with
a high accuracy we obtained new results for the $2s$ HFS interval.

Our optically measured results
\begin{equation}
f_{\rm HFS}^{\rm H}(2s) = 177\,556\,860(16)\;{\rm
Hz}\;,~~~\cite{H2s},
\end{equation}
\begin{equation}
f_{\rm HFS}^{\rm D}(2s) = 40\,924\,454(7)\;{\rm
Hz}\;,~~~\cite{D2s},
\end{equation}
agree with early microwave data and are somewhat more precise.

\section{The HFS tests of bound state QED: the summary}

\begin{table*}[htb]
\caption{Comparison of experiment and theory of hyperfine
structure in light hydrogen-like atoms. The experimental
references can be found in \cite{PRep1}. Here $\Delta$ is a
deviation of theory from experiment and $\sigma$ is a combined
uncertainty.\label{T:exp}} \label{table:2}
\newcommand{\m}{\hphantom{$-$}}
\newcommand{\cc}[1]{\multicolumn{1}{c}{#1}}
\renewcommand{\tabcolsep}{2.5pc} 
\renewcommand{\arraystretch}{1.2} 
\begin{tabular}{@{}lllll}
\hline
Atom, quantity & Experiment, [kHz] & Theory, [kHz] & $\Delta/\sigma$  \\
 \hline
{Mu, $1s$ HFS} & 4\,463\,302.78(5) & 4\,463\,302.88(55)& -0.18 \\
Ps, $1s$ HFS & 203\,389\,100(740)  & 203\,391\,700(500) & -2.9\\
Ps, $1s$ HFS & 203\,397\,500(1600) & & -2.5 \\
{H, $D_{21}$}  & 49.13(13) & 48.953(3)  & 1.4 \\
H,  $D_{21}$  &  48.53(23) &  & -1.8  \\
H,  $D_{21}$  &  49.13(40) & & 0.4  \\
{D, $D_{21}$}  & 11.280(56) & 11.312\,5(5) & -0.58  \\
D, $D_{21}$  &  11.16(16) &  & -1.0 \\
{$^3$He$^+$, $D_{21}$}~~~  &-1\,189.979(71) &-1\,190.08(15) &0.6 \\
$^3$He$^+$, $D_{21}$ & -1\,190.1(16) &  &  0.0  \\
\hline
\end{tabular}\\[2pt]
\end{table*}

We summarize state-of-the-art in the precision tests of the bound
state QED theory of the hyperfine structure in Table~\ref{T:exp}.
The theoretical accuracy is limited by our ability to calculate
higher-order radiative, recoil and radiative recoil effects (see
review \cite{PRep1} for more detail). The higher-order
contributions crucial for the uncertainty are related to the same
diagrams and thus all tests listed in the table are really
competitive. Theory and experiment are generally in good
agreement. There is a minor discrepancy for positronium up to
approximately 3 standard deviations, but statistically that is
acceptable if the tests as a whole are considered.

\section*{Acknowledgement}

The work was supported in part by DFG and RFBR (under grant \#
06-02-16156).

\end{document}